\documentclass[aps,prl,twocolumn,floatfix,superscriptaddress]{revtex4}

\usepackage{amssymb,amsmath,amstext}                
\usepackage{graphicx}
\usepackage{epstopdf}                                               
\usepackage{color}                                                     
\usepackage{bm}                                                        
\usepackage{appendix}

\usepackage[utf8]{inputenc}
\usepackage{lipsum}

\usepackage{ulem}
\usepackage{latexsym}
\usepackage[colorlinks=true,citecolor=blue,linkcolor=magenta]{hyperref}
\usepackage{soul}
\usepackage{amsmath}
\usepackage{float}

\def\be{\begin{equation}}
\def\ee{\end{equation}}
\def\bea{\begin{eqnarray}}
\def\eea{\end{eqnarray}}

\def\bi{\begin{itemize}}
\def\ei{\end{itemize}}

\definecolor{dgreen} {RGB}{78,138,21}

\definecolor{xuedong} {RGB}{0,128,0}

\definecolor{giergiel} {RGB}{0,168,128}

\definecolor{purple} {RGB}{128,0,160}

\begin{document}
\title{A Decade of Time Crystals: Quo Vadis?}
\author{Peter Hannaford}
\affiliation{
Optical Sciences Centre, Swinburne University of Technology, Hawthorn, Victoria 3122, Australia}
\author{Krzysztof Sacha}
\affiliation{
Instytut Fizyki Teoretycznej,
Uniwersytet Jagiello\'nski, ulica Profesora Stanis\l{}awa \L{}ojasiewicza 11, PL-30-348 Krak\'ow, Poland}

\begin{abstract}
Ten years ago, the new era of time crystals began. Time crystals are systems that behave in the time dimension like ordinary space crystals do in space dimensions. We present a brief history of a decade of research on time crystals, describe current research directions, indicate challenges, and discuss some future perspectives for condensed matter physics in the time domain.
\end{abstract}

\maketitle

{\bf Ten years of history of time crystals.} --
The new era of time crystals began in 2012 when Frank Wilczek alone and together with Alfred Shapere published two seminal papers on quantum and classical time crystals \cite{Wilczek2012,Shapere2012}. It is very well known that interacting quantum particles can self-organize their distribution in space and spontaneously form space crystals. However, prior to 2012 no one raised the question of whether similar self-organization could also happen in time. Wilczek's idea of quantum time crystals was ambitious because he proposed that such a crystal could spontaneously form even if a quantum many-body system were in the ground state. Soon after, it turned out that for many-body systems with quite general two-body interactions this was not feasible \cite{Bruno2013b,Watanabe2015,Syrwid2017}. 

The classical version of time crystals proposed by Shapere and Wilczek relied on a single-particle system with kinetic energy that possessed a minimum at non-zero velocity \cite{Shapere2012}. At the lowest possible energy, the particle performed periodic motion, and thus revealed crystalline structure in time. In the Hamilton description, this seemed impossible because a minimum of the classical Hamiltonian implied vanishing velocity. Shapere and Wilczek solved this puzzle by showing that at a minimum, the Hamiltonian had a cusp where the Hamilton equations were not well defined \cite{Shapere2012}. 

Let us return to quantum time crystals. The original Wilczek idea did not work but three years later there was a resurrection of quantum time crystals in closed but periodically driven systems (i.e. systems that are described by unitary evolution, and thus there is no coupling to the environment, but the energy is not a constant of motion because the Hamiltonian depends explicitly on time). In 2015 it was shown that ultra-cold atoms bouncing on an oscillating atom mirror could spontaneously self-reorganize their motion and start evolving with a period twice longer than the period dictated by the drive \cite{Sacha2015} --- a new periodic motion appeared, i.e., a novel crystalline structure emerged in the time domain. The following year, the same phenomena were proposed in spin-based solid state systems \cite{Khemani16,ElseFTC}. The spins were being periodically flipped and if there were interactions between them, the systems evolved with a period twice as long as the driving period even when the spin flipping process was imperfect. Periodically driven systems possess discrete time translation symmetry which corresponds to a translation in time by the period of the external drive. The symmetry is spontaneously broken when time crystals form and therefore they have been dubbed discrete (or Floquet) time crystals \cite{Yao2017}.

The first experiments on discrete time crystals were performed in 2017. Two groups implemented periodically driven spin systems and demonstrated that the presence of the spin-spin interactions resulted in periodic evolution of the systems with a period twice or three times as long as the period of the drive \cite{Zhang2017,Choi2017}. Period-doubling time crystals were later realized in other spin-based experiments \cite{Pal2018,Rovny2018,Kyprianidis2021,Randall2021,Mi2022,Frey2022} and in ultra-cold atoms \cite{Smits2018}. 

Discrete time crystals that evolve with a period twice as long as the period of the drive are small crystals in the time dimension. However, it was shown that big discrete time crystals evolving with a period even 100 times longer than the driving period are also attainable in the laboratory \cite{Giergiel2018a,Surace2018, Pizzi2019a,Giergiel2020}. These systems consist of many elementary cells in time and allow for realization of condensed matter phenomena in the time domain like in the case of ordinary space crystals, which consist of many elementary cells in space \cite{Hannaford2020PW}. It was also shown that spontaneous breaking of discrete time translation symmetry can lead to the formation of not only crystal structures in time, but also quasi-crystal structures in time \cite{Li2012,Giergiel2018c}. 

Apart from the described discrete time crystals in closed systems, other research directions of the field of time crystals have been developed, which we list in the following. 

In 2019 it turned out that in time-independent systems if, instead of two-body interactions, multi-body interactions were considered, it was possible to construct a system where time crystal behavior could emerge in the ground state \cite{Kozin2019}. How to realize such a system experimentally, which would be a realization of Wilczek's original idea of a quantum time crystal, is not known yet.

All time crystals described so far correspond to closed systems. However, it is known from the systematic studies of Prigogine and others that open systems that are out of thermodynamic equilibrium can possess steady states that spontaneously break the time translation symmetry and reveal periodic evolution in time \cite{Prigogine1977,PrigogineNobel}. In the context of time crystals, dissipative time-independent and periodically driven systems have recently been investigated theoretically and experimentally \cite{Gong2017,Iemini2017,Buca2019a,Dogra2019,Cosme2019,Kessler2020,Taheri2020}. They are important from the point of view of practical applications because in the real world, i.e., outside advanced laboratories, it can be difficult to avoid contact with the environment. 

If bosons form a Bose-Einstein condensate and the system is described in the framework of the grand canonical ensemble, then the number of particles fluctuates and only the average particle number can be well defined. Then, Bose–Einstein condensation is identified with breaking of the U(1) symmetry and the phase of the order parameter oscillates periodically in time, indicating that the time translation symmetry is spontaneously broken. The oscillations can be observed because the system is not isolated and can exchange particles with another system that plays the role of a reference frame with respect to which the oscillations can be measured \cite{Volovik2013}. Such spontaneous emergence of time periodic oscillations can be investigated in a condensate formed by bosonic magnetic quasi-particles (magnons) excited in superfluid \textsuperscript{3}He-B \cite{Volovik2013} (see also \cite{Homann2020,Ojeda2021}). Application of a radio-frequency pulse deflects the magnetization of the superfluid \textsuperscript{3}He-B and the resulting induction signal first decays but afterwards spontaneously reappears with its own phase completely forgetting the phase of the radio-frequency pulse. Spontaneous precession of the magnetization in magnon condensates in the context of time crystals has been experimentally investigated \cite{Autti2018,Kreil2018,Autti2021}. Note that when the total number of bosons is well defined, a Bose-Einstein condensate is identified with a single non-vanishing eigenvalue of the reduced single-particle density matrix and the U(1) symmetry broken approach cannot be used to define a time crystal.

Classical time crystals proposed by Shapere and Wilczek are time-independent systems that can perform periodic motion even if their energy is the lowest possible \cite{Shapere2012,Shapere2019,Das2018}. However, the research on quantum discrete time crystals also inspired investigation of classical discrete time crystals. The latter are periodically driven classical many-body systems which break ergodicity and perform motion with a period longer than the period of the drive \cite{Kim2006,Heo2010,Yao2018,Simula2022}. 

We have sketched a brief history of the new era of time crystals. More details can be found in review articles \cite{Sacha2017rev,else2019discrete,khemani2019brief,Guo2020} and in recent books \cite{SachaTC2020,GuoBook2021}.
\newline

{\bf Current research directions.} --
The existence of discrete time crystals in closed systems is surprising because a generic periodically driven quantum many-body system is normally expected to heat up to an infinite temperature state where no regular temporal structure can be observed. There is currently intensive research concerning the lack of heating and hence the ergodicity breaking by discrete time crystals in closed systems, which we will describe shortly.

Discrete time crystals in spin-based or lattice systems can be roughly divided into two groups: (i) prethermal time crystals and (ii) time crystals protected by many-body localization. In group (i) there are systems which are driven with high frequency $\omega$ so that to absorb a single quantum portion of energy from the time-periodic drive, a large number of local rearrangements in the system is needed \cite{Abanin2017a,Else17prx,Mizuta2019,Machado2019,khemani2019brief}. Consequently, they do not heat up for a long time and only thermalize at a time that increases exponentially with $\omega/J_{local}$, where the frequency $J_{local}$ is associated with the local interaction energy in the system. In the group (ii), there are systems whose static counterparts reveal many-body localization. Time-independent quantum many-body systems in the presence of disorder can be many-body localized. That is, they do not thermalize (in the sense of the eigenstate thermalization hypothesis \cite{Deutsch1991,srednicki94}) because there are local integrals of motion and they exhibit emergent integrability \cite{Huse14,Rahul15,Mierzejewski2018}. When a time-periodic driving is turned on and a discrete time crystal forms, the crystal will, in principle, live forever because the ergodicity breaking is protected by the many-body localization \cite{Khemani16,ElseFTC}. While such a mechanism is plausible, one should bear in mind that there is no rigorous proof that it is true in the thermodynamic limit and for an infinite evolution time. Even in the case of time-independent systems, many-body localization is a highly non-trivial phenomenon and there is only a class of one-dimensional time-independent systems where its existence is mathematically proven \cite{Imbrie2016}. It is very difficult to verify ergodicity breaking in the laboratory because in all experiments performed so far there were other processes, not included in the time crystal Hamiltonians, which made the lifetime of the crystals finite (typically a hundred of the driving periods \cite{Zhang2017,Choi2017,Pal2018,Rovny2018,Smits2018,
Kyprianidis2021,Randall2021,Mi2022,Frey2022}). However, prethermal properties of discrete time crystals not protected by many-body localization have been demonstrated experimentally \cite{Kyprianidis2021}. 

The first discrete time crystal was proposed in ultra-cold bosonic atoms bouncing on an oscillating atom mirror, i.e., in a  system which is very different from the spin-based or lattice systems \cite{Sacha2015}. First, the single-particle version of this system, contrary to a single spin, possesses a well-defined classical limit. Second, this is a bosonic many-body system. The lifetime of an ordinary space crystal is infinite if first the thermodynamic limit is taken, and only then the infinite evolution time is considered. Otherwise, a sufficiently long quantum evolution of a space crystal results in spreading of the center of mass position of the system and consequently blurring of the crystalline structure in space \cite{SachaTC2020}. In the case of ultra-cold bosonic atoms bouncing on a mirror, if the limit of an infinite number of the particles is taken, the description of a discrete time crystal reduces to the mean-field Gross-Pitaevskii equation \cite{Sacha2015,Wang2020,Wang2021}. Period-doubling solutions of the Gross-Pitaevskii equation are stable and consequently the discrete time crystal does not decay \cite{Sacha2015}.

The periodic evolution of discrete time crystals in closed systems implies that the average transfer of energy from the drive to the systems is zero. For these systems, the environment is an enemy which can destroy the time crystals \cite{Lazarides15}. In the case of dissipative discrete time crystals the environment participates in the formation of the crystals \cite{Gong2017}. The balance between energy pumped by an external drive and energy released to the environment is responsible for the formation of periodically evolving steady states. Treating the environment as an ally is thus another strategy to realize time crystals, which can be used in future practical applications. Two experiments demonstrating dissipative discrete time crystals have been performed recently, in ultra-cold atoms \cite{Kessler2020} and in an optical system \cite{Taheri2020}. In the latter case, one can realize big discrete time crystals that can live forever, i.e., as long as the optical resonator is pumped by laser beams. Moreover, the experiment was carried out at room temperature and does not require advanced experimental equipment, and consequently, transfer from the laboratory to the real world seems straightforward. 

One may ask the question why do we need time crystals? They may be interesting from the point of view of basic research, but what are their practical applications? Discrete time crystals are many-body systems that can maintain coherent evolution for a long time. This is a very important property which can be used in future applications. Whether this will allow one to build quantum computers, we shall see. Will time crystals allow us to build precise clocks? At least in the case of discrete time crystals, the precision of their periodic evolution is limited by the precision of the external periodic drive. However, the fact that they evolve with subharmonic frequencies can be used as a frequency division, which in the optical frequency regime, like in the experiment \cite{Taheri2020}, is highly non-trivial. In order to think about practical applications of time crystals, they must be easy to implement and well controlled experimentally. Therefore, we await the development of experimental techniques and new groundbreaking realizations of time crystals.

All discrete time crystals realized in spin-based systems and ultra-cold atoms so far have been small in the time dimension \cite{Zhang2017,Choi2017,Pal2018,Rovny2018,Smits2018,Kyprianidis2021,Randall2021,Kessler2020,Mi2022,Frey2022}. Typically, they evolved with a period twice as long as the driving period. In other words, they consisted of only two elementary cells in time. With such time crystals it is not possible to do condensed matter physics in the time domain. However, discrete time crystals that are big in the time dimension can be realized in ultra-cold atoms \cite{Giergiel2018a,Giergiel2020,Pizzi2019a} 
and also in optical systems \cite{Taheri2020}. Especially the former system can host a variety of solid-state phases which we are going to describe in the following \cite{SachaTC2020,GuoBook2021}. The development of solid-state physics in the time dimension is a natural next stage of time crystal research with powerful potential for future practical applications. This research which is in the stage of experimental implementation \cite{Giergiel2020} is described below.
\newline

{\bf Condensed Matter Physics in Big Time Crystals.} -- 
Discrete time crystals created by a Bose-Einstein condensate (BEC) of ultracold atoms bouncing on an oscillating atom mirror can exhibit dramatic breaking of discrete time translational symmetry, where the bouncing atoms can evolve with periods up to about 100 times longer than the period of the driving mirror \cite{Giergiel2018a,Giergiel2020}. Such a system allows the creation of big time crystals possessing a large number of temporal lattice sites and are suitable for investigating condensed matter physics in the time dimension \cite{Sacha15a,sacha16,delande17,Matus2019,Mierzejewski2017,Giergiel2018b,Giergiel2018,
Giergiel2018c,Kosior2018,Giergiel2021,Kuros2021}. These condensed matter phenomena may or may not involve spontaneous breaking of the discrete time translation symmetry. In the latter case, the formation of time lattices resembles the situation of photonic time crystals that do not arise spontaneously because they are externally imposed by periodic fabrication of the refractive index in space \cite{Joannopoulos_Book}.

For ultra-cold atoms bouncing on an oscillating mirror, the time-periodic driving function of the mirror can be expanded as a Fourier series $f(t)=f(t+2\pi/\omega)=\sum_k f_k e^{ik\omega t}$, which by a proper choice of the Fourier components $f_k$ allows one to engineer effective temporal lattice potentials with nearly any shape. This provides a highly flexible platform to realize a broad range of condensed matter phenomena in the time dimension. 
We first consider the case of a single particle bouncing on a mirror oscillating with period $T$. The classical effective Hamiltonian describing the motion in the vicinity of a resonant trajectory evolving with period $sT$ (where $s$ is integer), in the moving frame of the resonant trajectory, has the form \cite{SachaTC2020} 
\be
H_{eff}=\frac{P^{2}}{2m_{eff}} + V_{eff} \left( \Theta \right), 
\label{singheff}
\ee
where we use action-angle coordinates $(I,\theta)$ and all quantities are in gravitational units \cite{Sacha2015}. $P=I-I_s$, $I_s$ is the action for the $s$ resonant trajectory, $\Theta=\theta-\omega t/s$, $m_{eff}$ is an effective mass and $V_{eff}(\Theta)$ is the effective potential which depends on the Fourier components $f_k$ of the time-periodic driving function $f(t)$ of the mirror. For pure harmonic driving in time, $f(t)\propto\cos\omega t$, the effective potential $V_{eff}\propto \cos(s\Theta)$. Then, in the quantum description of the particle \cite{SachaTC2020,Giergiel2020}, the eigen-energies form energy bands and the corresponding eigenstates have the form of Bloch waves in time when they are observed in the laboratory frame. If we are interested in just the first energy band, the description reduces to the tight-binding model \cite{Dutta2015} in which there are $s$ Wannier-like states $w_j(z,t)$ corresponding to localized wave-packets evolving along the resonant trajectory with period $sT$, in a similar way as for the problem of an electron in a periodic space crystal. 

For the {\it many-body} case of $N$ indistinguishable bosonic particles, if the interaction energy per particle is much smaller than the energy gap between the first and second energy bands of the corresponding single-particle problem, we may restrict attention to the first energy band. The many-body Floquet Hamiltonian then takes the form of an effective many-body Bose-Hubbard Hamiltonian which describes $N$ bosons in a time lattice \cite{SachaTC2020}
\be
\hat H_F
\approx -\frac{J}{2}\sum_{j=1}^s\left(\hat a_{j+1}^\dagger\hat a_j+H.c.\right)+\frac12 \sum_{i,j=1}^sU_{ij}\hat a_i^\dagger \hat a_j^\dagger \hat a_j \hat a_i, 
\label{manybh}
\ee
where the bosonic annihilation operators $\hat a_i$ correspond to periodically evolving Wannier-like wave-packets $w_j(z,t)$. $J$ is the tunnelling amplitude of the particle between neighbouring wave-packets, and $U_{ii}$ and $U_{ij}$ are the intra-site and inter-site interaction coefficients, which are proportional to the interparticle contact interaction $g_0(t)$, and hence to the s-wave scattering length, which can be modulated in time by modulating the magnetic field of a Feshbach resonance.   

In space crystals we are interested in the regular distribution of particles in space at a fixed moment of time  (i.e., at the detection moment). In time crystals, the roles of space and time are exchanged, i.e., we fix a position in space and ask if the probability of clicking of a detector located at this point behaves periodically in time \cite{SachaTC2020}. The Hamiltonian (\ref{manybh}) describes a many-body system loaded into a time lattice. The detector located close to the resonant trajectory of this system will reveal a crystalline structure in time, and the underlying temporal behavior is identical to the spatial behavior of a solid state system described by the static version of the Hamiltonian (\ref{manybh}), i.e., when the operators $\hat a_i$ correspond to time-independent Wannier modes of a space lattice \cite{Dutta2015}. In the following we present a few examples of non-trivial condensed matter phenomena which can be realized in the time domain with the help of ultra-cold atoms bouncing on a mirror \cite{Hannaford2020PW,SachaTC2020}.

Condensed matter phenomena can also be investigated in photonic time crystals which are temporal analogues of photonic space crystals but the refractive index of the dielectric material changes periodically in time and not in space \cite{Lustig2018,Sharabi2021,Galiffi2022}. Another direction of condensed matter research in periodically driven systems is phase-space crystals \cite{Guo2013,Guo2016,Guo2016a,Liang2017,
Guo2020,Guo2021,GuoBook2021}. They reveal crystal structures in phase space where single-particle and many-body solid state phenomena can be investigated.
\newline

\begin{figure}[t]
\centering
\includegraphics[width=0.238\textwidth]{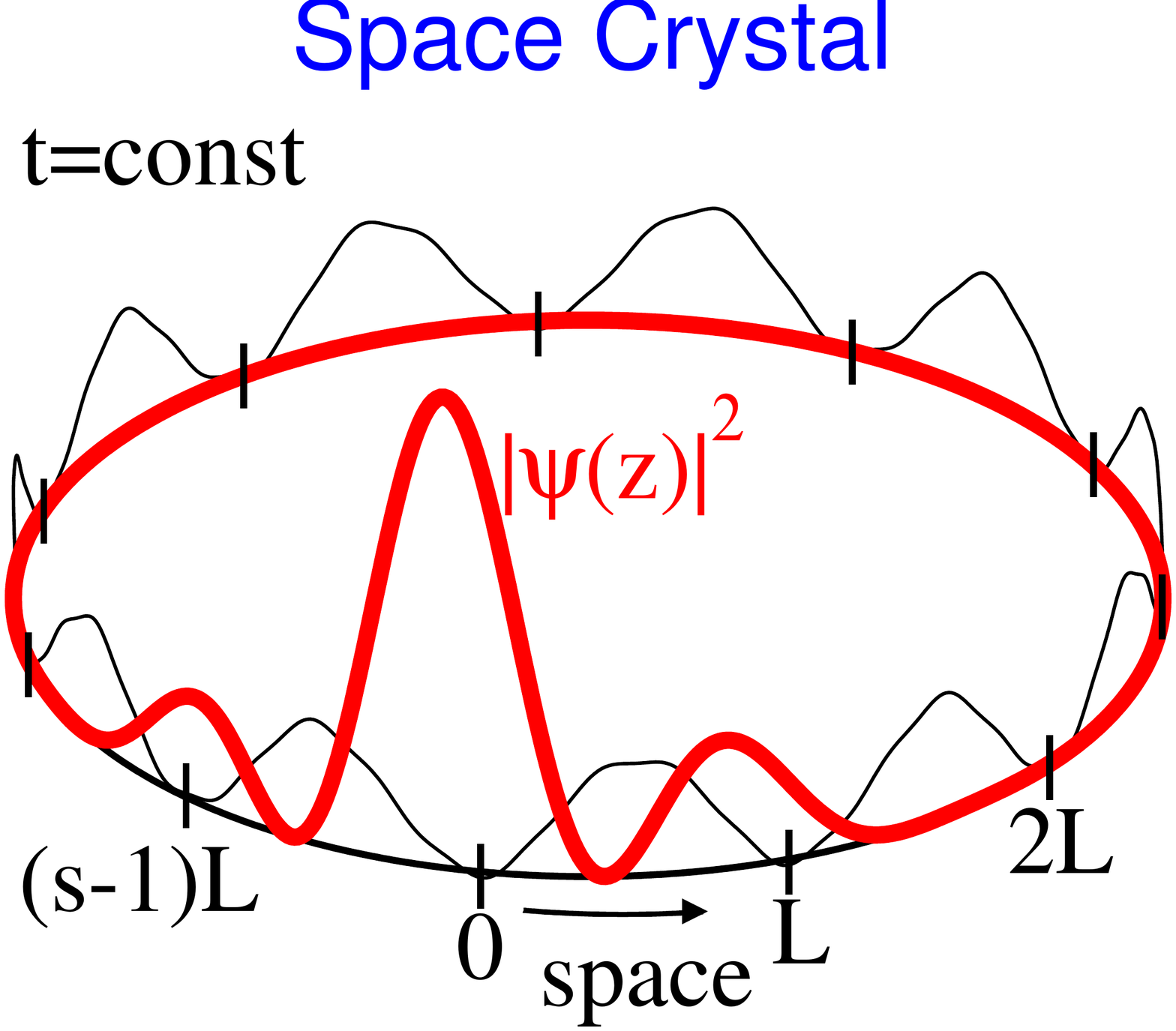}
\includegraphics[width=0.238\textwidth]{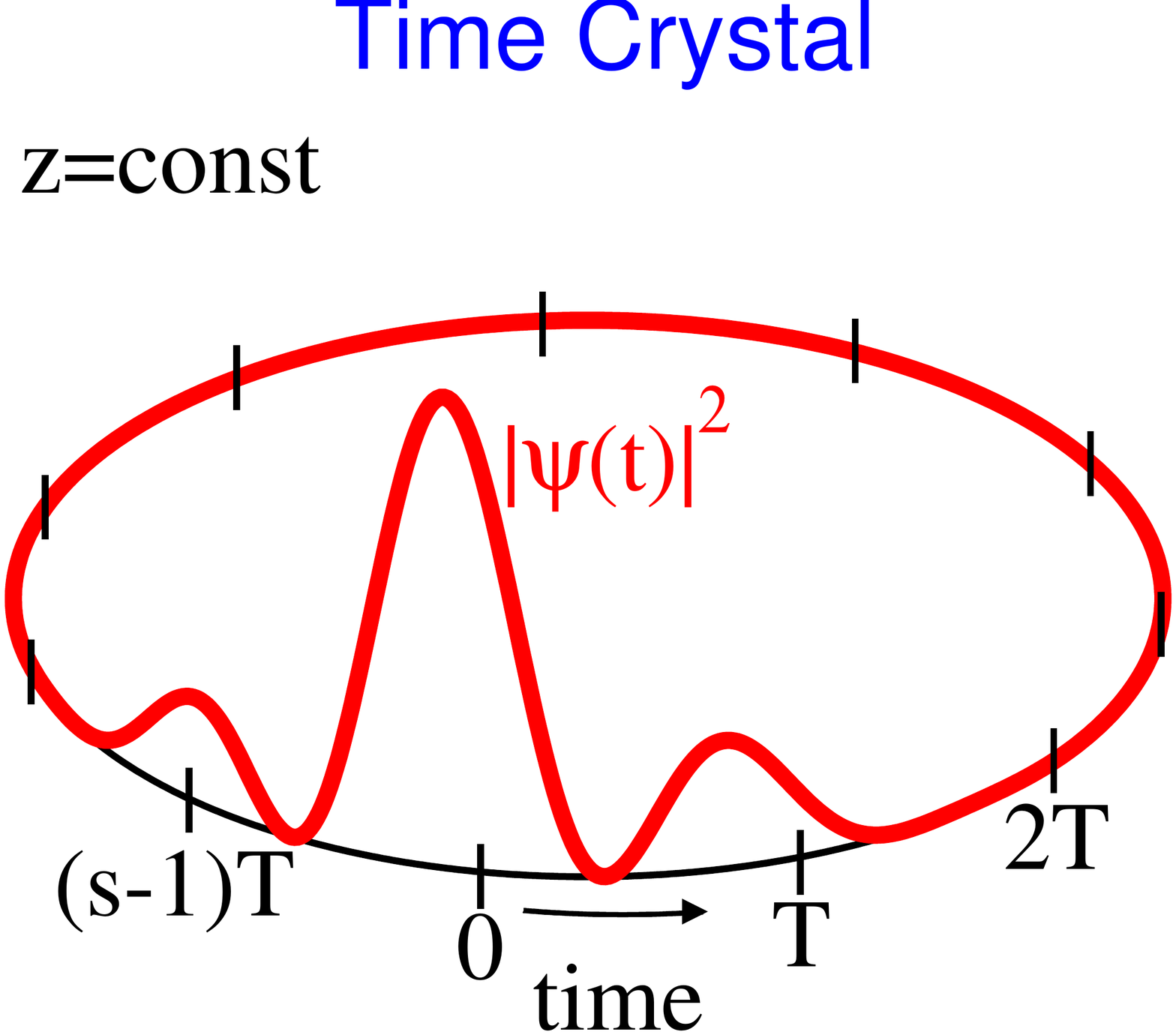}
\caption{Comparison of Anderson localization in a 1D space crystal with periodic boundary conditions (left) with Anderson localization in the time domain (right). A particle moving in the space crystal with disorder can Anderson localize, i.e., if we go around the ring, we observe an exponential localization of a particle around a certain point in space. When switching from Anderson localization in space to Anderson localization in time, we have to exchange the role of space and time. We fix the position in space and ask if the probability of clicking of a detector is exponentially localized around a certain moment of time --- such behavior is repeated periodically due to the periodic boundary conditions in time. This circulating ring system is equivalent to an atom bouncing resonantly on an oscillating mirror. Reprinted from \cite{Sacha2017rev}.}
\label{singAnder}
\end{figure}

{\it Anderson localization in the time dimension.} 
Anderson localization normally involves the exponential localization of eigenstates of a particle in space due to the presence of a spatially disordered potential \cite{Anderson1958}. Anderson localization can also be observed in the time domain due to the presence of temporal disorder \cite{Sacha15a}.

If a \emph{single} \emph{particle} 
is bouncing resonantly on a mirror which is driven in time as $f(t) = \lambda\cos(\omega t) + \sum_{k}f_{k}e^{ik\omega t/s}$, where the $f_{k}$ are random complex numbers with $|f_{k}| \ll \lambda$, there are additional temporal disorder terms, $\sum_{j = 1}^{s}\epsilon_{j}|a_{j}|^{2}$ (where the $\epsilon_{j}$ are real random numbers), in the tight-binding Hamiltonian [i.e., the single-particle version of the Hamiltonian (\ref{manybh})], and we observe an exponential localization in time, i.e., the probability of clicking of a detector located close to the resonant trajectory is exponentially localized around a certain moment of time \cite{Sacha15a,Giergiel2018a}, see Fig. 1. It is also possible to observe the localized-delocalized Anderson transition in the time domain if a particle moving in 3D space is properly driven \cite{delande17}. Recently, signatures of temporal Anderson localization have also been studied in photonic time crystals \cite{Sharabi2021}.

If a resonantly driven \emph{many-body} system experiences temporal disorder, there are additional temporal disorder terms, $\sum_{j=1}^s\epsilon_{j}\hat a_j^\dagger\hat a_j$ in the effective Bose-Hubbard model (\ref{manybh}), and the system can reveal many-body localization in time \cite{Mierzejewski2017}. 
\newline

{\it Topological time crystals.}
Topological insulators are insulators in the bulk but possess topologically protected conducting edge states or surface states.

\begin{figure}
\centering
\includegraphics[width=0.48\textwidth]{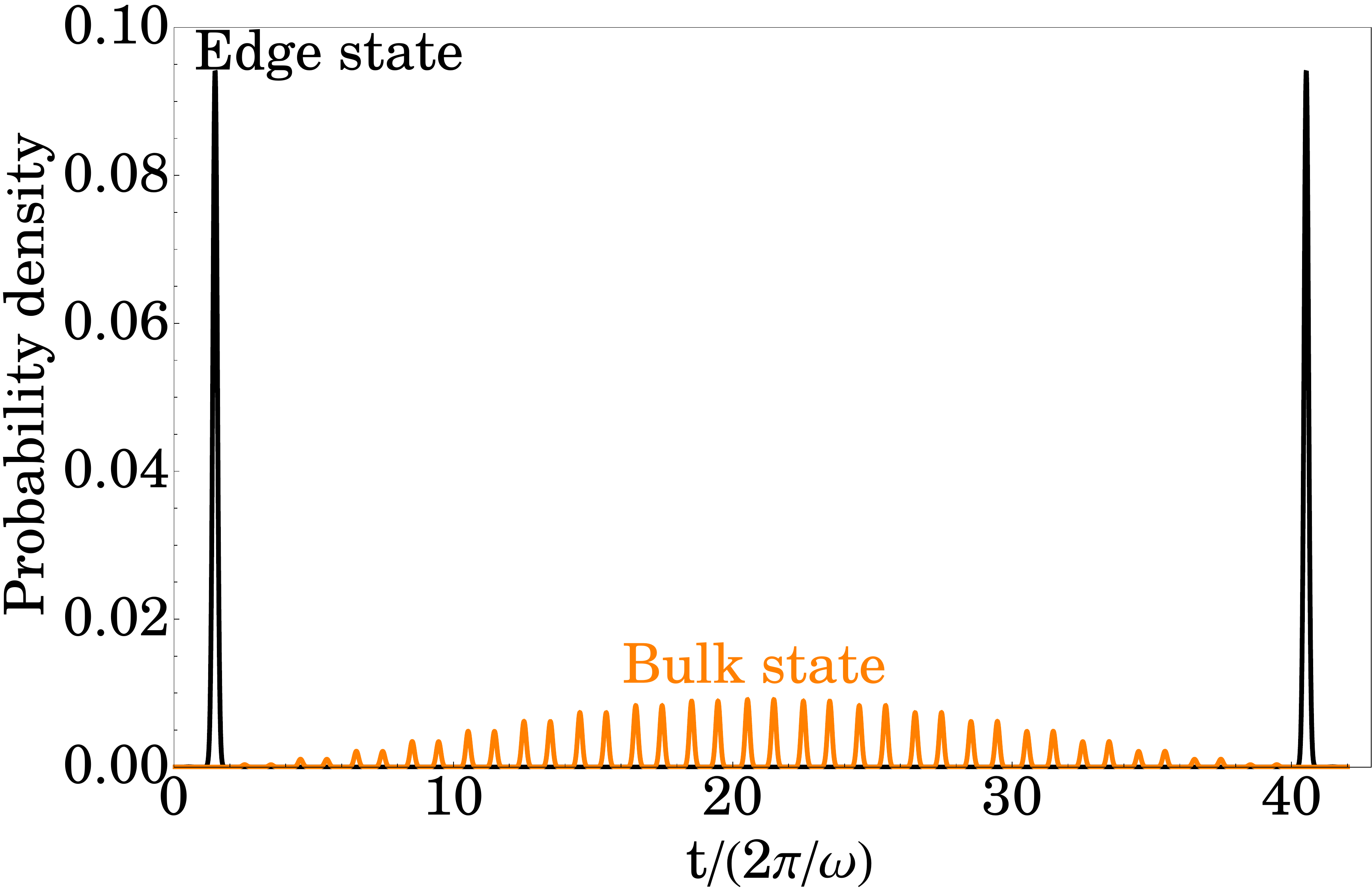}
\caption{A particle driven resonantly in time so that the effective Hamiltonian (\ref{singheff}) reproduces the Su-Schrieffer-Heeger model \cite{Su1979} with an edge for the case $s = 42$. In the laboratory frame, the bulk states are delocalized along the entire resonant orbit while the topologically protected edge states are localized close to the edge created in time. Reprinted from \cite{SachaTC2020}.}  \label{fig_edge} 
\end{figure}

If a \emph{single particle} is resonantly bouncing on a mirror which is driven with two harmonics, $f(t) = \lambda\cos(s\omega t) + \lambda'\cos(s\omega t/2)$, then the effective potential in Eq.~(\ref{singheff}) reveals a crystalline structure with a two-point basis $V_{eff}(\Theta) = V_{0}\cos(s\Theta) + V_{0}'\cos(s\Theta/2)$. In the quantum description, the first two energy bands of $H_{eff}$ can be described by the Su-Schriefer-Heeger (SSH) model which is a tight-binding model with staggered hopping amplitudes \cite{Su1979}. By introducing a narrow barrier in $V_{eff}(\Theta)$, by applying a properly chosen additional modulation $\sum_{k}^{}f_{k}e^{ik\omega t}$, we can create a topological system with an edge and two eigenstates which are localized close to the edge and appear localized in time when the particle is observed in the laboratory frame, see Fig.~\ref{fig_edge} \cite{Giergiel2018b}. 

Lusting {\it et al.} have considered topological photonic time crystals \cite{Lustig2018}. If an electromagnetic wave propagates in a time-varying medium so that two photonic time crystals are realized one by one, and if the wavenumber of the wave is located in a common gap of the crystals with different topological properties, then a temporal analogue of a topologically protected edge state can be observed.

In a \emph{many-body} system, if the interaction $g_{0}$ is modulated not only in time but also in space, via a Feshbach resonance, there are additional possibilities to engineer different effective long-range interactions. For example, if $g_{0}\left( z,t \right) = \sum_{m = 2}^{2}{\alpha_{m}(t)z^{m}}$, the coefficients $\alpha_{m}(t)$ can be chosen so that the Bose-Hubbard model (\ref{manybh}) reduces to a form which can reveal the bosonic analogue of the topological Haldane insulator in a spin-one chain \cite{Giergiel2018b}.
\newline

{\it Mott insulator phases in the time dimension.}
If in a resonantly driven $N$-atom system, Eq.~(\ref{manybh}), the effective repulsive on-site interactions $U_{ii}$ dominate over the long-range repulsion $U_{ij \neq i}$ and are sufficiently strong compared with the tunneling rate $J$, the system reveals a Mott insulator phase in the time domain \cite{Sacha15a}. Then, the gap between the ground and excited states of the Bose-Hubbard Hamiltonian (\ref{manybh}) is opened, fluctuations of the number of atoms in each site are suppressed, and the system is not compressible. When we look at a fixed position close to the resonant trajectory, well-defined bunches of atoms (with number $N/s$) arrive periodically at the observation point like on a conveyer belt and there is no coherence between them \cite{Sacha15a}. This is in contrast to the superfluid phase where the repulsive interactions are weak and the ground state of the Bose-Hubbard model (\ref{manybh}) is a BEC. The superfluid-Mott insulator transition can be realized in the time lattice described here by changing the tunneling rate $J$ \cite{Sacha15a}.
\newline
 
{\it Exotic long-range interactions in the time dimension.}
If we periodically modulate the contact interaction $g_{0}(t)$ between ultra-cold atoms by modulating the magnetic field around a Feshbach resonance, we can make the (normally negligible) long-range interactions $U_{ij\ne i}$ in the effective Hamiltonian (\ref{manybh}) significant and engineer how they change with distance $|i - j|$ between temporal lattice sites \cite{Giergiel2018}. If $g_{0}(t)$ changes in time with the period $sT$, one can engineer exotic long-range interactions including interactions not existing in nature, such as interactions that change the repulsive and attractive character in an oscillatory way with increasing $|i - j|$. 
\newline 

\begin{figure}
\centering
\includegraphics[width=0.35\textwidth]{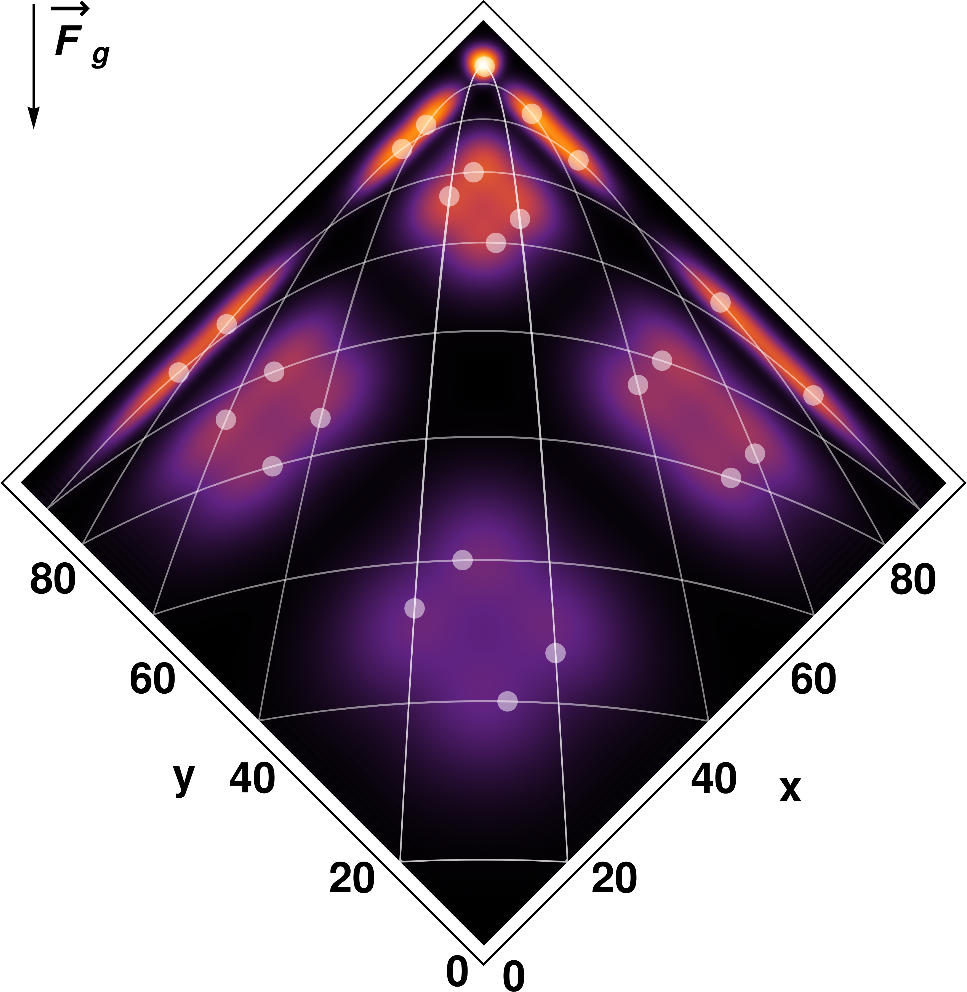}
\caption{Trajectories of atomic wave-packets bouncing between two orthogonal
oscillating mirrors, located at $x = 0$ and $y = 0$, on a 2D $5 \times 5$ time lattice. Reprinted from \cite{Giergiel2018}.}
\label{fig_multi}
\end{figure}

{\it Higher-dimensional time lattices.} 
Time crystals can be extended to 2D (or 3D) by allowing the BEC to bounce resonantly between two (or three) oscillating atom mirrors \cite{Giergiel2018}.

We first consider the case of two \emph{orthogonal} oscillating mirrors, located at $x = 0$ and $y = 0$, oscillating at frequency $\omega$ and inclined at 45\textsuperscript{o} to the gravitational force (Fig.~\ref{fig_multi}). With this configuration, the localized wave-packets are products of 1D wave-packets. If the $s$ resonance condition is fulfilled for both mirrors, then a square $s \times s$ time lattice is created (Fig.~\ref{fig_multi}). Use of this 2D time lattice allows gradual breaking of the discrete time translation symmetry when the motion along one of the two orthogonal directions breaks the symmetry and with a range of system parameters the motion along the other direction also reveals symmetry breaking \cite{Kuros2021}.

For atoms bouncing between a pair of orthogonal atom mirrors, it is possible to realize spontaneous formation of \emph{time quasi-crystals} when the ratio $s_{x}/s_{y}$ (where $s_{x,y}$ are the resonance numbers corresponding to the particle's motion along the $x$ and $y$ directions) approximates the golden number $(1+\sqrt{5})/2 \approx 1.618$ \cite{Giergiel2018c}. Then, when the discrete time translation symmetry is spontaneously broken, the sequence of atoms reflected from one mirror and the other mirror forms a Fibonacci quasi-crystal in time.

\begin{figure}
\centering
\includegraphics[width=0.49\textwidth]{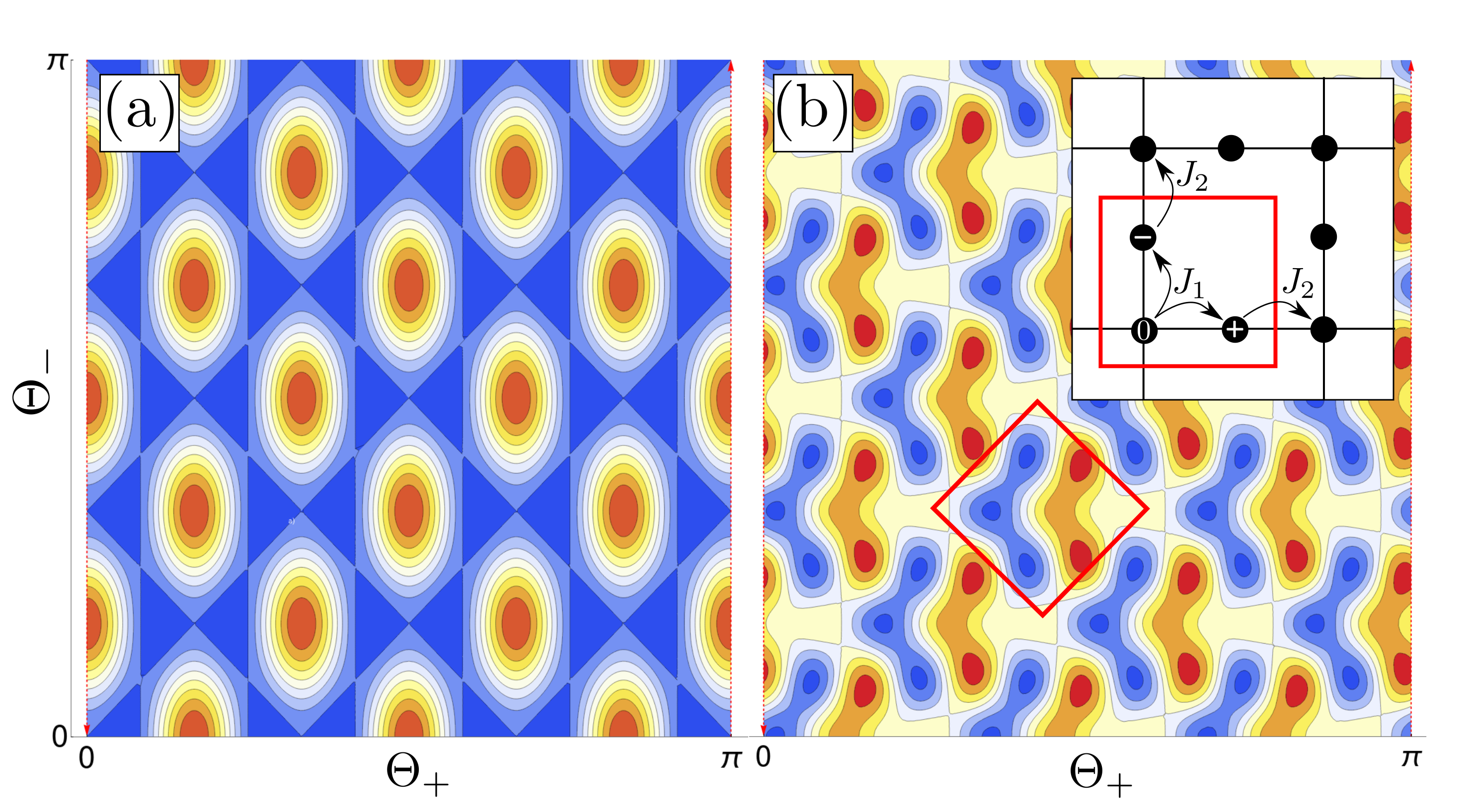}
\caption{(a) Honeycomb and (b) Lieb time lattices presented in the frame moving along a resonant trajectory. Atoms loaded to lattices fulfill the boundary conditions of the M\"obius strip geometry, i.e., points $\{\Theta_{+} = \pi,\Theta_{-}\}$ are identified with $\{\Theta_{+} = 0,\pi - \Theta_{-}\}$. Reprinted from \cite{Giergiel2021}.}
\label{fig_mobius}
\end{figure}

We now consider the case when the oscillating mirror at $y = 0$ is moved to the vertical position to form a 45\textsuperscript{o} wedge. When a particle strikes the vertical mirror, its momenta are exchanged $p_x \leftrightarrow p_y$, whereas when it strikes the 45\textsuperscript{o} mirror, the momentum $p_y$ remains the same while the momentum $p_x$ is reversed $p_x \rightarrow-p_x$ \cite{Giergiel2021}. With this configuration, the system is inseparable and forms a time lattice with a {\it M\"obius strip} geometry. Using different relative amplitudes and phases of the oscillations of the two mirrors we can realize different crystalline structures of the effective potential, such as a honeycomb time lattice or a Lieb square time lattice (Fig.~\ref{fig_mobius}) with a `flat band', in which the dynamics of the atoms is governed solely by interactions \cite{Giergiel2021}.

{\bf Summary.} -- 
Seminal papers published by Wilczek and Shapere in 2012 initiated the new era of time crystals. The idea of time crystals is inspiring scientists and attracting public attention. We have briefly described the history of time crystals and pointed out current research directions. Among them, condensed matter physics in the time domain seems to be a promising horizon for novel theoretical and experimental research. It is also a natural next stage of development of time crystals toward practical applications. 

\acknowledgments
This work was supported by the Australian Research Council Discovery Grant No. DP190100815 and the National Science Centre, Poland via Project No.~218/31/B/ST2/00319.


\end{document}